\documentclass[twocolumn,showpacs,prb]{revtex4}

\usepackage[dvips]{graphicx}% Include figure files
\usepackage{dcolumn}% Align table columns on decimal point
\usepackage{bm}% bold math

\begin{document}

\title{Alkali-metal-induced Fermi level and two dimensional electrons\\ at cleaved InAs(110) surfaces}

\author{Masaaki Minowa, Ryuichi Masutomi, Toshimitsu Mochizuki, and Tohru Okamoto}
\affiliation{Department of Physics, University of Tokyo, 7-3-1 Hongo, Bunkyo-ku, Tokyo 113-0033, Japan}

\date{June 3, 2008}

\begin{abstract}
Low-temperature Hall measurements have been performed on two-dimensional electron gases (2DEGs) induced by deposition of Cs or Na on {\it in situ} cleaved surfaces of $p$-type InAs.
The surface donor level, at which the Fermi energy of the 2DEG is pinned, is calculated from the observed saturation surface electron density using a surface potential determined self-consistently.
The results are compared to those of previous photoelectron spectroscopy measurements.

\end{abstract}

\maketitle

It is well known that a two-dimensional electron gas (2DEG) can be easily formed on the surface of narrow band-gap III-V semiconductors.
Photoelectron spectroscopy measurements have shown that the position of the Fermi level lies above the conduction-band minimum at cleaved (110) surfaces of InAs and InSb with various kinds of adsorbed materials. \cite{Baier1986, Aristov1991, Aristov1993, AristovInSb1993, Aristov1994, Nowak1995, Aristov1995, Morgenstern2000, Getzlaff2001, Betti2001}
Recent observations of the quantum Hall effect reveal the perfect two dimensionality of the surface inversion layers formed on $p$-type substrates. \cite{Tsuji2005,Masutomi2007}
In Ref.~\onlinecite{Tsuji2005}, the conduction electron density $N_s$ at the surface of $p$-InAs was obtained from the Hall coefficient as a function of the coverage of adsorbed Ag.
The saturation of $N_s$ with increasing Ag atomic density was explained in terms of a pinning of the Fermi level $\varepsilon_F$ of the 2DEG at an adsorbate-induced surface donor level.
In this work, we extend the experiment to 2DEGs on InAs(110) surfaces induced by submonolayer deposition of alkali metals.
The obtained $\varepsilon_F$ pinning position is plotted as a function of the atomic ionization energy \cite{Monch1988} and compared to the data from previous photoelectron spectroscopy measurements.

The samples used were cut from a Zn-doped single crystal with an acceptor concentration of 1.2 $\times$ $10^{17}$ cm$^{-3}$.
Sample preparation and experimental procedures are similar to those used in Ref.~ \onlinecite{Tsuji2005}.
Two current electrodes and four voltage electrodes were prepared by deposition of gold films onto noncleaved surfaces at room temperature.
Sample cleaving, subsequent deposition of adsorbates, and Hall measurements were performed at low temperatures in an ultra-high vacuum chamber cooled down to liquid ${}^4$He temperature. 
Before the deposition of Cs or Na on the cleaved surface ($3~{\rm mm} \times 0.4~{\rm mm}$), resistance between any two electrodes was greater than 400~k$\Omega$ at 4.2~K.
Alkali metals were deposited from aluminosilicate ion sources. \cite{Weber1966,Feeney1976,Ong1994}
The atomic density $N_{\rm ad}$ of the adsorbates was obtained by integrating the ion current using a setup illustrated in Fig.~1.
The Hall measurements were made at 2.0~K in a magnetic field applied perpendicular to the optically flat cleaved surface.
Since the density and mobility of the 2DEG do not change when the sample is left overnight at 2.0 K, we believe that the base pressure is low enough and contamination effects are negligible.
\begin{figure}[b!]
\includegraphics[width=5cm]{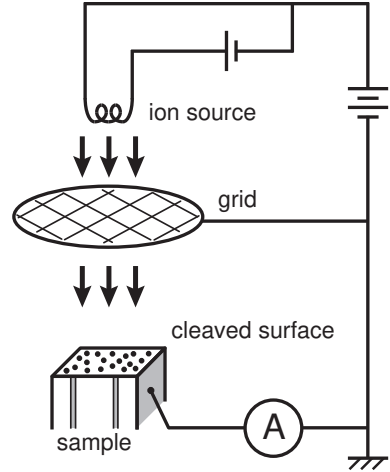}
\caption{
Schematic view of the experimental setup mounted in a low-temperature vacuum chamber.
Alkali metals are deposited from the aluminosilicate ion source (Refs.~\onlinecite{Weber1966,Feeney1976,Ong1994}) which is well out gassed before  {\it in situ} cleaving.
The grid is used to reduce the electric field near the sample which may lead to non-uniform distribution of the adsorbates. 
}
\end{figure}

Figure 2 shows the conduction electron density $N_s$ obtained from the Hall coefficient.
The results for the depositions of Cs and Na are plotted as a function of $N_{\rm ad}$ together with the data for Ag-deposition reported in Ref.~\onlinecite{Tsuji2005}.
In the low $N_{\rm ad}$ region, $N_s$ increases with a slope of approximately equal to unity.
This indicates that each adsorbed atom donates one electron to the inversion layer after the completion of the depletion layer formation. \cite{Tsuji2005}
The saturation value of $N_s$ depends on the kind of adsorbates.
This can be attributed to the difference in the surface donor level at which $\varepsilon_F$ of the 2DEG is pinned. \cite{Tsuji2005}

\begin{figure}[t!]
\includegraphics[width=7.5cm]{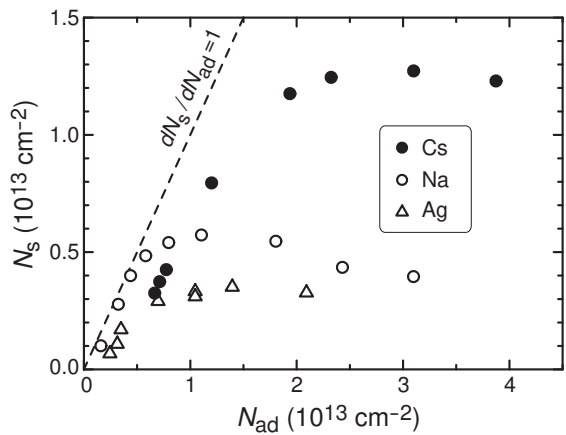}
\caption{
Conduction electron density in the inversion layer as a function of the surface atomic density of adsorbates.
Data for Ag are taken from Ref.~\onlinecite{Tsuji2005}.
The dashed straight line indicates a slope of unity.
}
\end{figure}

In order to obtain the $\varepsilon_F$ positions for the saturation values of $N_s$, we calculate subband energies and densities in the inversion layer.
Our procedure is similar to that proposed by Ando, \cite{Ando1985} except that the surface potential is determined self-consistently and the depletion layer charge is taken into account.
The nonparabolic bulk dispersion relation used is expressed as
\begin{eqnarray}
\varepsilon_k=-\frac{1}{2}\varepsilon_g+\left[ \frac{1}{4}\varepsilon_g^2+\varepsilon_g \frac{\hbar^2 k^2}{2m^\ast} \right]^{1/2}
,
\end{eqnarray}
where $\mathbf{k}=(k_x,k_y,k_z)$ is the three-dimensional wave vector, $\varepsilon_g$ is the band gap, and $m^\ast$ is the effective mass at the conduction band minimum.
This relation is rewritten as
\begin{eqnarray}
\frac{\hbar^2 k^2}{2m^\ast} =\varepsilon_k \left( 1+\frac{\varepsilon_k}{\varepsilon_g} \right)
.
\end{eqnarray}
In spite of its simplicity, Eq.~(1) is in good agreement with the recent elaborate calculation using the $30 \times 30$ $\mathbf{k \cdot p}$ approach. \cite{Radhia2007}
The deviation is less than 10~\% or 50~meV up to $\varepsilon_k=1000$~meV if $\varepsilon_g=417$~meV and $m^\ast=0.026 m_e$ \cite{Vurgaftman2001} are used.
In a triangular-like surface potential $V(z)$,
the energy $\varepsilon_n(k_\parallel)$ of the subband $n$ ($=0,1,{\mathrm {etc.}}$) and the in-plane wave vector $k_\parallel=(k_x,k_y)$ is given in the WKB approximation by
\begin{eqnarray}
\int^{z_n(k_\parallel)}_0 k_z dz=\left(n+\frac{3}{4}\right) \pi,
\end{eqnarray}
with
\begin{eqnarray}
\frac{\hbar^2}{2m^\ast}(k_z^2+k_\parallel^2)
=[ \varepsilon_n(k_\parallel)-V(z) ] \left[1+ \frac{\varepsilon_n(k_\parallel)-V(z)}{\varepsilon_g}\right],
\end{eqnarray}
where we take the origin of the $z$-axis at the surface and $z_n(k_\parallel)$ is the turning point defined by
$k_z=0$. \cite{Ando1985}
The Fermi wave vector $k_{F n}$ of the subband $n$ is obtained as the solution of Eq.~(3) by putting $k_\parallel=k_{F n}$ and $\varepsilon_n(k_\parallel)=\varepsilon_F$ in Eq.~(4).
The inversion layer electron density $N_s$ is the sum of the subband electron density $k_{Fn}^2/2\pi$.

The surface potential $V(z)$ is related to the charge distribution by Poisson's equation.
In contrast to the classical approximation adopted in the simplified scheme in Ref.~\onlinecite{Ando1985}, we determine $V(z)$ from the subband electron distribution self-consistently.
The envelope function $\zeta_n(z)$ of the subband $n$ is approximated by the Airy function, which is well known as the eigenfunction in the triangular potential well, as
\begin{eqnarray}
\zeta_n(z)=c_n {\mathrm {Ai}} \{ \alpha_n [z/z_n(k_\parallel)-1] \},
\end{eqnarray}
where $c_n$ is the normalization factor and $-\alpha_n$ is the $n$th zero of the Airy function [${\mathrm {Ai}}(-\alpha_n)=0, n=0,1,{\mathrm {etc.}}$].
The electron distribution is obtained from the sum of $| \zeta_n(z) |^2$ for all $n$ and $k_\parallel$.
\begin{figure}[t!]
\includegraphics[width=7.5cm]{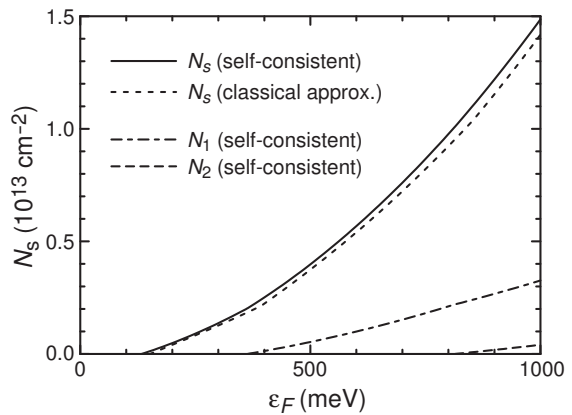}
\caption{
Calculated $N_s$ vs Fermi energy $\varepsilon_F$ measured from the conduction band minimum at the surface.
The solid line is calculated using $V(z)$ determined self-consistently.
The results obtained from the classical approximation for $V(z)$ (Ref.~\onlinecite{Ando1985}) are also plotted as the dotted line.
Depletion layer charges arising from ionized acceptors were taken into account and their potential was replaced by a constant electric field. \cite{Ando1982}
$N_1$ and $N_2$ are the subband electron densities while $N_0=N_s-N_1-N_2$ is not shown explicitly.
}
\end{figure}
In Fig.~3, the results of the self-consistent calculation \cite{check} are shown together with those obtained from the classical approximation for $V(z)$. \cite{Ando1985}
Electrons occupy the 1st (2nd) excited subband above $\varepsilon_F =363$~meV (814~meV).
The multi-subband structure was confirmed by a Fourier transform analysis of the Shubnikov-de Haas oscillations observed in the longitudinal magnetoresistance.\cite{Darr1978}

Using the experimental data in Fig.~2 and the calculated relationship between $N_s$ and $\varepsilon_F$ in Fig.~3, the maximum $\varepsilon_F$ positions are obtained for Cs, Na, and Ag. \cite{AgCorrection}
In Fig.~4, the results are plotted on the graph first given by Aristov {\it et al.} who performed photoelectron spectroscopy measurements. \cite{Aristov1994}
In the case of GaAs(110) surfaces, a tight-binding calculation of the donor-type surface state energy for $s$-electron adatoms is in good agreement with the experimental data which show a negative dependence on the atomic ionization energy. \cite{Monch1988}
A similar tendency can be seen in the data for InAs(110) surfaces including the present results.
However, our data are significantly higher than those obtained from previous photoelectron spectroscopy measurements for Cs and Na.
This may be partly due to the difference in the deposition temperature.
In Ref.~\onlinecite{Morgenstern2000}, the bond formation of Fe adatoms observed by a scanning tunneling microscope was related to thermal diffusion at 300~K.
For an InSb surface covered with 0.8~monolayer Ag, a drastic decrease in $N_s$ with increasing annealing temperature was observed above 15~K. \cite{Masutomi2007}
It is expected that thermal diffusion drives the formation of dimers, chains, clusters, etc. and drastically reduces the number of isolated adatoms.

\begin{figure}[t!]
\includegraphics[width=7.5cm]{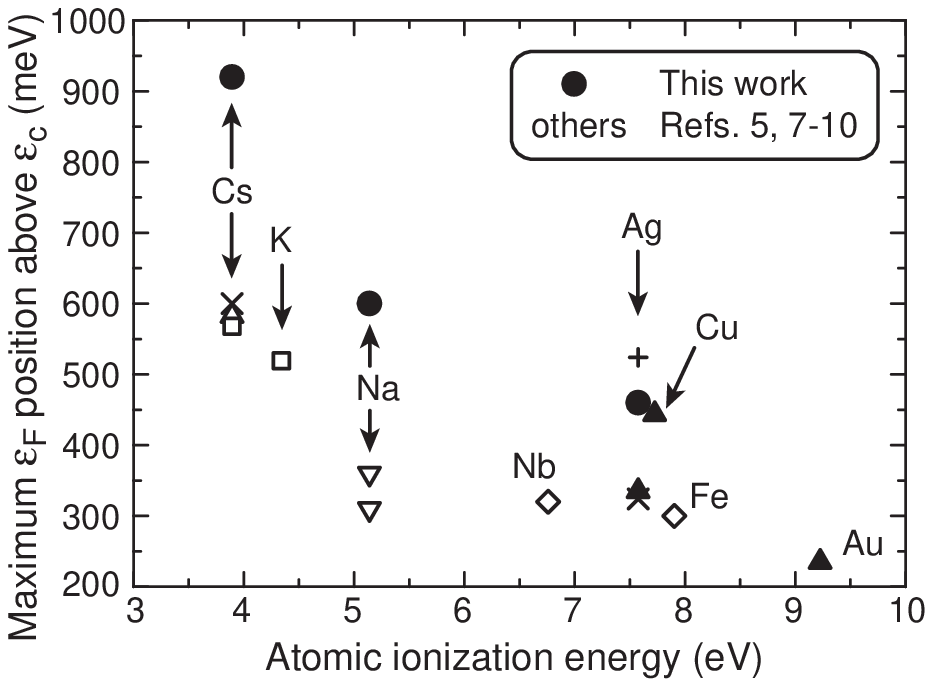}
\caption{
The maximum $\varepsilon_F$ position measured from the conduction band minimum at the surface as a function of the first ionization energy of the corresponding atom.
The open and filled symbols indicate depositions at room and low (4.2-20~K) temperatures, respectively: circles, this work; triangles, Ref.~\onlinecite{Aristov1994}; inverted triangles, Ref.~\onlinecite{Aristov1995}; diamonds, Refs.~\onlinecite{Morgenstern2000} and \onlinecite{Getzlaff2001}; squares, Ref.~\onlinecite{Betti2001}.
Data obtained at 80~K (plus) and 200~K (crosses) in Ref.~\onlinecite{Betti2001} are also shown.
}
\end{figure}

In summary, we have studied the evolution of the surface electron density $N_s$ with depositions of alkali-metals at liquid-helium temperatures.
The positions of the Fermi energy for the saturation values of $N_s$ were deduced from a self-consistent calculation and compared to those obtained from photoelectron spectroscopy measurements.

This work was partly supported by Grant-in-Aid for Scientific Research (B) (Grant No. 18340080), Grant-in-Aid for Scientific Research on Priority Area "Physics of new quantum phases in superclean materials" (Grant No. 18043008), and Grant-in-Aid for JSPS Foundation (Grant No. 1811418) from MEXT, Japan.


\begin{thebibliography}{30}
\bibitem{check} The subband energy obtained from the approximation using Eq.~(5) was compared to the exact self-consistent calculation for the case when the bulk dispersion relation is parabolic and only the lowest subband is occupied. It is lower by less than 10~\% for various ratios of $N_s$ to the depletion layer charge density.
\end{thebibliography}
\end{document}